# Active vase mirrors warped by Zernike polynomials for Correcting off-axis aberrations of fixed primary mirrors.
## 2 - Optical testing and performance evaluation

G. Moretto[1], G. R. Lemaître[2], T. Bactivelane[2], M. Wang[1], M. Ferrari[2], S. Mazzanti[2], B. Di Biagio[2], and E.F. Borra[1]

[1] Centre d'Optique, Photonique et Laser - COPL, Département de Physique - Université Laval
  Québec - QUE - Canada G1K7P4
[2] Observatoire de Marseille, Laboratoire d'Optique, Université de Provence, 2 Place Le Verrier
Marseille CEDEX- 13248 - France

**Abstract.** We investigate the aspherization of an active mirror for correcting third and fifth-order aberrations. We use a stainless steel AISI 420 mirror with a controlled pressure load, two series of 12-punctual radial positions of force application distributed symmetrically in two concentric rings around the mirror. We obtain the wavefronts for *Cv1, Sph3, Coma3, Astm3, Comatri, Astm5* aswell as those of the added wavefronts. Although this active prototype mirror has general uses, our goal is to compensate the aberrations of a liquid mirror observing at large angles from the zenith.

## 1. Introduction

It has been known for some time that a rotating container filled with a reflecting liquid could in principle be used as the primary mirror of a telescope; but the concept has only recently been seriously studied. Borra et al (1992) have tested a 1.5-m diameter liquid mirror, showing that it was diffraction limited. This was followed by a 2.5-m mirror (Borra, Content and Girard, 1993), of equally good optical quality. Liquid mirrors are also being used for research; for example a 2.65-m diameter liquid mirror telescope has been successfully operated in the winter of 1994 ( Hickson et al. 1994), and a 2.65-m liquid mirror has also been successfully operated for over 2 years at the University of Western Ontario as the receiver of a lidar system.

Since a liquid mirror cannot be tilted, it only can observe a small region of sky near the zenith. Its usefulness is obviously increased if corrector designs can be found that give it access to a larger region of the sky. Following an article by Richardson & Morbey (1987), Borra (1993) has explored analytically the fundamental limits within which one may correct the aberrations of a liquid mirror observing off-axis. This simple analysis reached the surprising conclusion that aberrations could, in principle, be corrected in small patches to zenith distances as high as 45 degrees. In a recent paper (Wang, Moretto, Borra and Lemaître, 1994) we have investigated the design of a practical one-mirror active corrector that uses the technology pioneered by Lemaître (1974, 1989) to correct the aberrations of a parabolic mirror observing at large angles from its optical axis. The optical design of the corrector is simple, for one applies the wavefront correction by warping a secondary mirror into complex shapes that remove the aberrations with a brute-force approach. Ray-tracing computations have shown that the system can yield subarcsecond images within 10-arcseconds sub-regions inside a 20-degree field. While these narrow fields are inefficient for imagery, they can be useful for applications such as single-object or area spectroscopy.

Wang, Moretto, Borra and Lemaître (1994) only carried out computer work, and the feasibility of generating the required complex surfaces still has to be demonstrated. In this paper, we investigate whether it is possible to apply wavefront corrections up to the third and fifth-order Seidel aberrations by warping a small active mirror. The warping of this active mirror is done by using air pressure to add spherical aberration and defocus and by applying forces to 24 positions distributed symmetrically in two concentric rings around the mirror (Figures (3) & (2)) for the other aberrations. Details relative to the theory and elasticity design of this system were presented in part 1 of this article - *"Theory and Elasticity Design"* (Lemaître and Wang, 1994).

## 2. High-order aspherization surface

To produce an optical surface with higher-order aspherization of the test mirror, we use a similar terminology to that of the optical wavefronts, which in polar coordinates can be expressed as a linear combination of Zernike polynomials as

$$W = \sum_{nm} A_{nm} r^n \cos m\theta, \qquad (1)$$

where r and $\theta$ are the polar coordinates over the pupil. We consider the case of a symmetrically distributed load with respect to the center of a constant thickness test mirror.

A Fourier expansion into flexure modes k for the corrective elastic deflections W yields

$$W = \sum_{k=0}^{\infty} g_k w_k(r) \cos k\theta \equiv \sum_{k=0}^{\infty} z_k, \qquad (2)$$

where the modes k are orthogonal so that all modes $k \neq m$ do not interfere with the correction of a particular mode of order m.

From part 1 of this article, we already have the design of this active mirror from elasticity theory. The axial force distributions $F_{ik}$ applied to the internal ring, r = a and $F_{ek}$ applied to the external ring, r = c (where k= 1,2, ..., 12) are summarized in Table (1), for a Peak-to-Valley (PtV) deflection of 10 $\mu m$. This distribution of axial forces will provide the framework in which we will conduct the deformation of the active mirror.

## 3. Experimental set-up

The optical tests of the active mirror, were done with a Fizeau interferometer. In order to obtain fringes with good visibility we used the combination of a microscope objective and a helium-neon gas laser at 6328 $\mathring{A}$ operating in a single longitudinal mode. A schematic diagram is shown in Figure (1)

As shown in Figure (3) and Figure (2), at each end of the arms is mounted a $6 - 8\ mm$ differential screw connected to a base plate (A), that is three times thicker than the active mirror. One run of the screw produces an axial displacement of $\pm 250\ \mu m$. This base plate (A) is fixed onto another base (B) by three support columns, arranged symetrically. This base (B) is centrally affixed onto the base of the interferometer (C).

To produce the aspherization we axially warp the test-mirror. This is accomplished by applying a uniform load. A first method uses the pressure difference generated by the air pressure control inside the active mirror, as shown in Figure (3) & (2). This produces a combination of symmetrical terms $Cv1$ and $Sph3$. In addition one second method modifies the forces on the two series of 12-punctual radial support positions r = a or/and r = c. This is done by raising or lowering the nuts providing an upward or downward displacement. We used Table (1) as a reference to know what displacement was necessary. Using the three screws at the base (B) of the test mirror we adjust this mirror until the image reflected from it comes into coincidence with the pinhole.

The beam splitter is placed to allow the CCD camera to observe the fringes. We used a $512 \times 582$ CCD array with a 20 $\mu m$ pixels, providing an image area of $6.4 \times 4.8\ mm^2$.

## 4. Results

The pure modes that were imposed on the active mirror are presented in Table (2), where we can see the 1st order mode $Cv1$, the 3rd order modes $Sph3$, $Coma3$, $Astm3$ and the 5th order modes $Comatri$ (triangular coma) and $Astm5$.

The first step is to put the mirror in the zero mode using a passive force distribution over r = a and r = c. This allowed us to check the optical quality of the mirror which was found to be 1.51$\lambda$ PtV, when in a stage of zero shell. After slightly optimizing the force distribution over the rings, which leads mainly to a small correction in $Sph3$, a plane mirror interferogram was obtained, as Figure (4) showing $\lambda/4$ PtV.

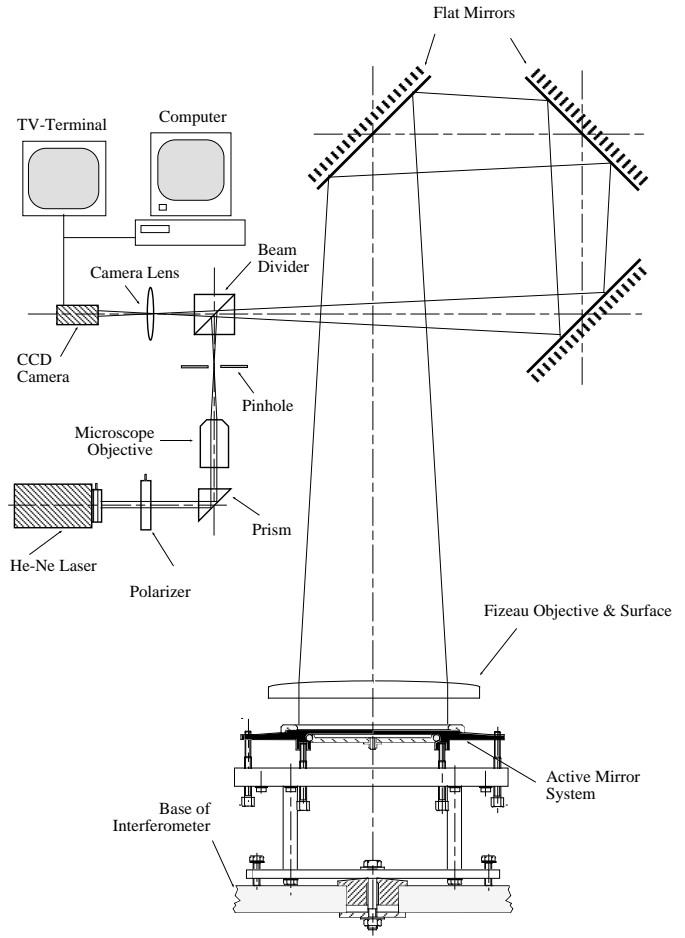

**Fig. 1.** Setup for the optical testing of the active mirror. It uses a Fizeau interferometer.

**Table 2.** Pure modes added to the active mirror.

| Mode k | Aberration | Term | Figure |
|---|---|---|---|
| k = 0 | $Cv1$ | $\rho^2$ | Figure (5a) |
| k = 0 | $Sph3$ | $\rho^4 + g_1 \rho^2$ | Figure (5b) |
| k = 1 | $Coma3$ | $\rho \cos \theta$ | Figure (6a) |
| k = 2 | $Astm3$ | $\rho^2 \cos 2\theta$ | Figure (6b) |
| k = 3 | $Comatri$ | $\rho^3 \cos 3\theta$ | Figure (6c) |
| k = 2 | $Astm5$ | $(\rho^4 + g_2 \rho^2) \cos 2\theta$ | Figure (7a,b) |

The pure curvature term $Cv1$ (k = 0) is generated by a uniform moment $M_a$ at the mirror contour r = a. To achieve, this we apply a uniform distribution of forces by having $F_a + F_c = 0$. This force distribution produces the interferometric pattern displayed by Figure (5a). The term $Sph3$ (k = 0) is obtained by generating air pressure or depressure under the mirror. If the equilibrium is only achieved by $F_a$, i.e $F_c = 0$,

Table 1. The axially distributed forces at the 24 discrete positions of the active mirror for a 10 $\mu$m deformation (PtV) are given below for each aberration mode.

Units:daN

| Angle | Number | $Cv1$ | | $Sph3^a$ | | $Astm3$ | | $Coma3$ | | $Comatri$ | |
| | of points | n=2, m=0 | | n=4, m=0 | | n=2, m=2 | | n=3, m=1 | | n=3, m=3 | |
| $\theta$ | N | Fa | Fc | Fa | Fc | Fa | Fc | Fa | Fc | Fa | Fc |
| 0 | 1 | -1.685 | 1.685 | -7.624 | 4.469 | 4.143 | 1.907 | -1.425 | 1.823 | 15.639 | 7.855 |
| $\pi/6$ | 2 | -1.685 | 1.685 | -7.624 | 4.469 | 2.071 | 0.953 | -1.234 | 1.579 | 0 | 0 |
| $\pi/3$ | 3 | -1.685 | 1.685 | -7.624 | 4.469 | -2.071 | -0.953 | -0.712 | 0.911 | -15.639 | -7.855 |
| $\pi/2$ | 4 | -1.685 | 1.685 | -7.624 | 4.469 | -4.143 | -1.907 | 0 | 0 | 0 | 0 |
| $2\pi/3$ | 5 | -1.685 | 1.685 | -7.624 | 4.469 | -2.071 | -0.953 | 0.712 | -0.911 | 15.639 | 7.855 |
| $5\pi/6$ | 6 | -1.685 | 1.685 | -7.624 | 4.469 | 2.071 | 0.953 | 1.234 | -1.579 | 0 | 0 |
| $\pi$ | 7 | -1.685 | 1.685 | -7.624 | 4.469 | 4.143 | 1.907 | 1.425 | -1.823 | -15.639 | -7.855 |
| $7\pi/6$ | 8 | -1.685 | 1.685 | -7.624 | 4.469 | 2.071 | 0.953 | 1.234 | 1.579 | 0 | 0 |
| $4\pi/3$ | 9 | -1.685 | 1.685 | -7.624 | 4.469 | -2.071 | -0.953 | 0.712 | -0.912 | 15.639 | 7.855 |
| $3\pi/2$ | 10 | -1.685 | 1.685 | -7.624 | 4.469 | -4.143 | -1.907 | 0 | 0 | 0 | 0 |
| $5\pi/3$ | 11 | -1.685 | 1.685 | -7.624 | 4.469 | -2.071 | -0.953 | -0.712 | 0.911 | -15.639 | -7.855 |
| $11\pi/6$ | 12 | -1.685 | 1.685 | -7.624 | 4.469 | 2.071 | 0.953 | -1.234 | 1.579 | 0 | 0 |

$^a$The uniform load required for producing the $Sph3$ mode is $q = 0.1883\ daN/cm^2$

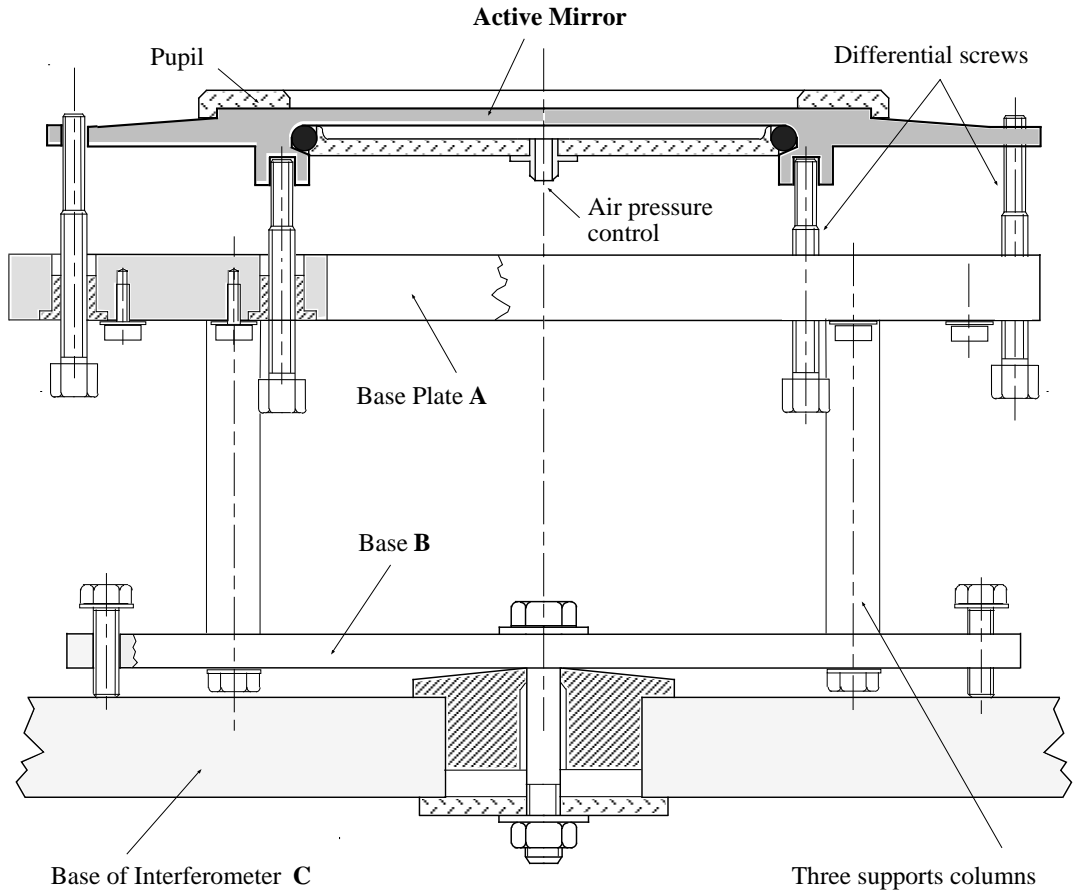

Fig. 2. The bases to constrain the active mirror

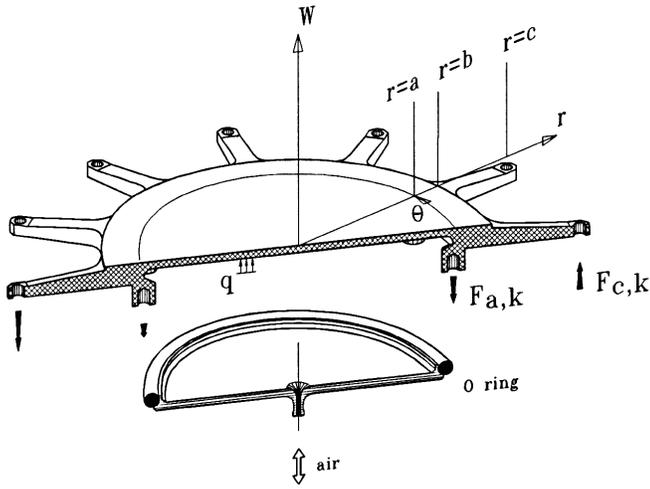

**Fig. 3.** The active mirror and its deformation system.

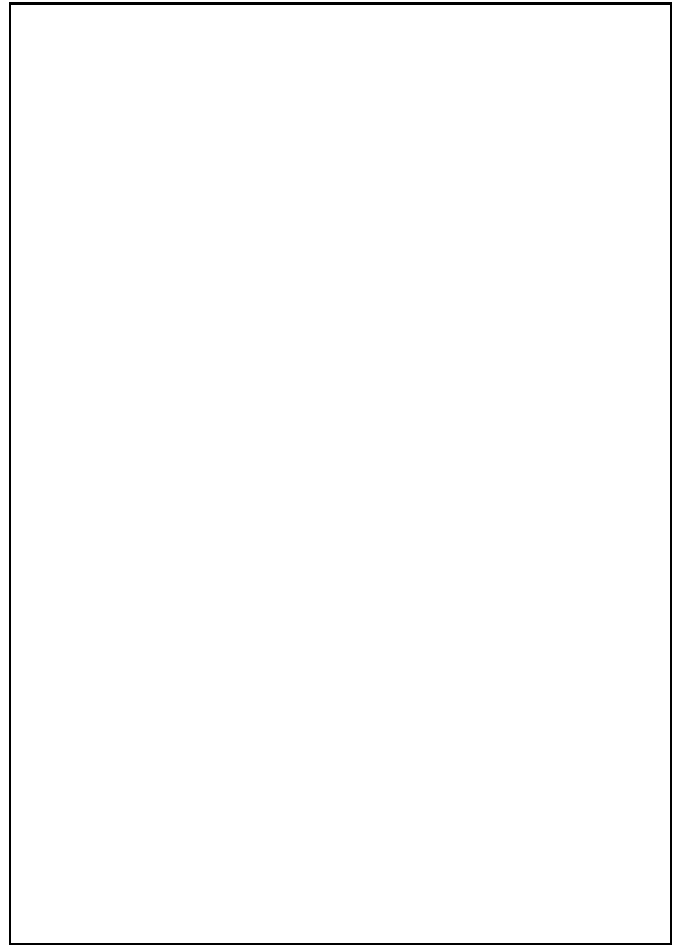

**Fig. 4.** Interferometric pattern of the active mirror in its initial stage , (a) without *Tilt1* and (b) with slight *Tilt1*.Tiny corrections has been applied to improve the optical figure issued from polishing.

the deformation is obtained with the combination of a *Cv1* mode as displayed by Figure (5a). A pure deformation at the 4th power of r is obtained by applying forces $F_a$ and $F_c$ such as shown in Table(1).

Pure modes *Coma3* ($k = 1$), *Astm3* ($k = 2$) and *Comatri* ($k = 3$) are generated by the modulation of forces at the 12-punctual radial support positions $r = a$ or/and $r = c$ with the distribution given in Table (1). When we used a specific mirror ring $r = a$ or $r = c$, we were obliged to optimize the forces exerted at the other contour support points. For example if we applied the forces to the mirror at $r = a$, we then optimized for zero forces around the mirror at $r = c$. This was carried out by visually checking the interferometric image for each loading mode. The Fizeau test used a convexo-plane lens having $F_{lens} = 2000\ mm$ and a flat surface at $\lambda/20 - PtV$.

The pure term *Coma3* with a $\cos\theta$ angular modulation, was mainly obtained by axial forces at $N = 1$ and 7. A slight tilt of the Fizeau's lens balances the deformation yielding to $Z = z_{11} + z_{31}$, whose deformation pattern is shown in Figure (6a).

*Astm3* with a $\cos 2\theta$ modulation was mainly produces by axial forces at azimuth numbers $N = 1, 4, 7$ and 10. The resulting interference pattern is displayed by Figure (6b).

The $\cos 3\theta$ modulation of *Comatri* was realized by forces mainly acting with the azimuth numbers $N = 1, 3, 5, 7, 9$ and 11. The resulting fringe pattern is displayed by Figure (6c).

*Astm5*, although not required for our present purpose, has been obtained by the addition of air pressure to the case of *Astm3*. Figure (7) display two deformation-type of this fifth order mode.

With regards to the superposition of *Cv1, Coma3, Sph3, Astm3* and *Comatri* modes, we must bear in mind that, as discussed above in equation (2), all modes $k \neq m$ are orthogonal and will not interfere with the correction of a particular mode of order m. It is assumed that accurate surface measurement is still possible with only a few fringes of low spatial frequency aberrations, i.e. astigmatism, coma and spherical aberration.

Figure (8a) shows a fringe pattern obtained from the superposition of *Coma3, Astm3* and a small tilt of the Fizeau-lens i.e a wavefront expressed by $Z = z_{11} + z_{22} + z_{31}$. To obtain this, we first generated *Coma3* as described before and followed with superposition of *Astm3*.

Next, *Comatri* is superposed onto the previous wavefront. The resulting fringe pattern is shown by Figure (8b).

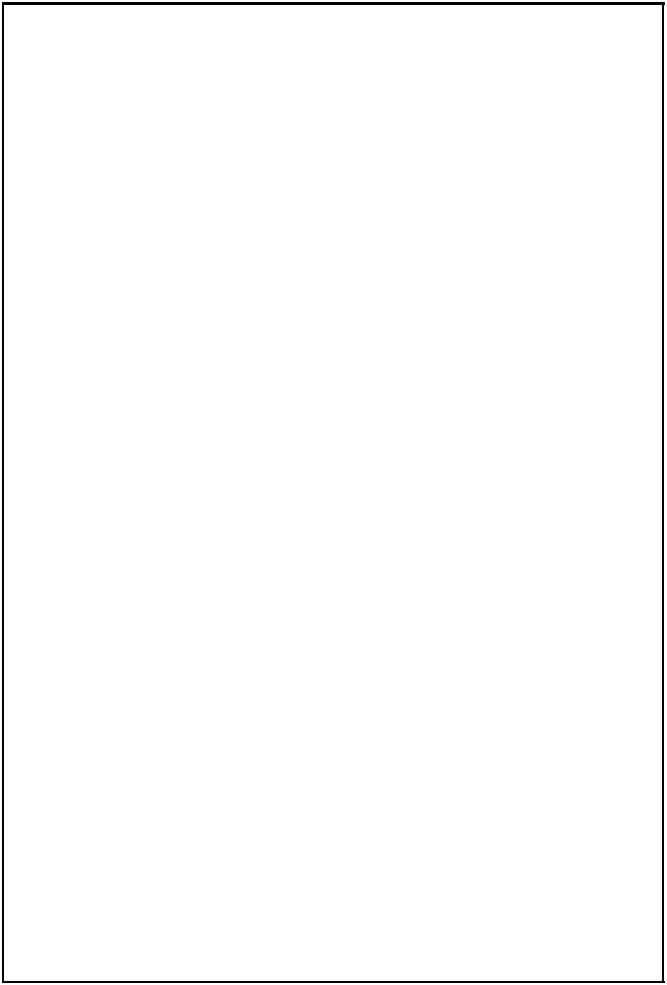

**Fig. 5.** Interferometric patterns of deformation modes (a) *Cv1* and (b) *Sph3*.

Finally, *Cv1* and *Sph3* modes were superposed onto the previous figure by a simple air pressure. The result is that the five correctable terms *(Cv1, Coma3, Astm3, Sph3* and *Comatri)*, make up a perfect "phantom shaped" interferometer pattern represented by $Z = z_{11} + z_{20} + z_{22} + z_{31} + z_{33} + z_{40}$. This corresponds to the final form of the active mirror surface necessary for the LMT, as shown in Figure (8c).

### 5. Fringe Analysis

The analysis of the fringes is performed using a combination of hardware and software package (FAST! V/AI) to digitalize the fringes of the interference pattern from the Fizeau interferometer. The digitized fringes are ordered and analyzed by the sofware that fits the Zernike polynomials to the wavefront.

The contour plot and isometric plot maps of the wavefront *Comatri* and the superposition wavefronts are shown in Figures (9a,b,c & d). Each contour plot is drawn with a fixed number of wave intervals and each interval is displayed in a different color. The isometric 3D plot is drawn at a fixed orientation and each color shows a different height of the surface. Table (3) shows the final aberration coefficients correspondig to Figures (9a, b, c & d).

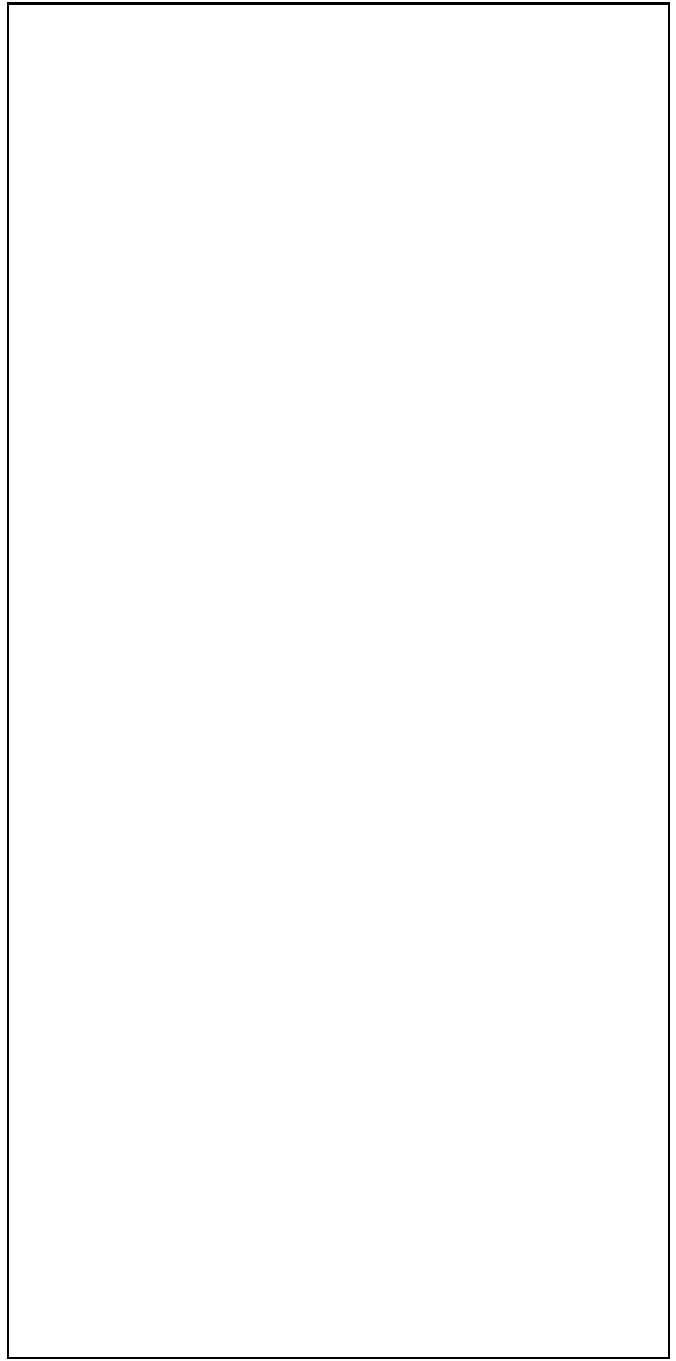

**Fig. 6.** Pure deformation modes (a) *Coma3*, (b) *Astm3* and (c) *Comatri*.

As we can see from Table (3), the generated *Coma3* term has introduced a mechanical tilt aberration and the air-pressured uniform loading on the mirror has brought about a negative *Sph3* aberration with some curvature variance. A compensation of the mechanical tilt would have to be done with the detection. A given moment $M_r$ applied around the mirror contour, produced by the modified forces on the ring r=c, will compensate the curvature variance.

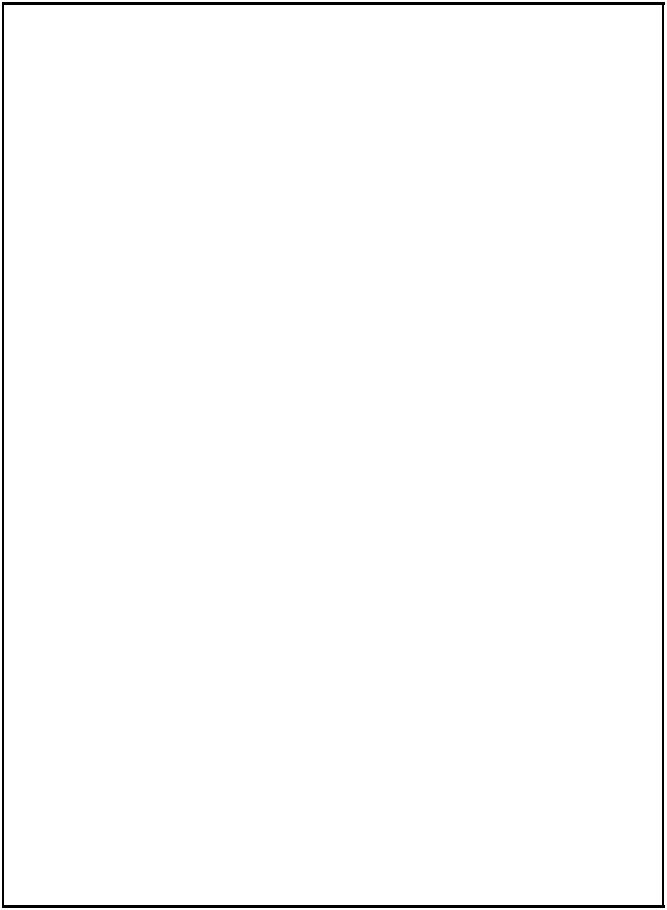

**Fig. 7.** Interferograms of (a) Astm5 bisymetric and (b) with a zero curvature mode in one direction.

## 6. Conclusion

We have investigated the feasibility of correcting third and fifth-order wavefront aberrations by warping an active mirror. In its final form, active warping of the mirror surface was achieved with, $Z = z_{11} + z_{20} + z_{22} + z_{31} + z_{33} + z_{40}$. Correction of $Cv1$, $Coma3$, $Astm3$, $Sph3$ and $Comatri$ was achieved via the simple procedure of warping a constant-thickness "fond de vase" active mirror with a pressure control and with forces applied to two series of 12 radial support positions distributed symmetrically in two concentric ring around the mirror. These are the type of corrections needed to make the corrector, designed by Wang, Moretto, Borra and Lemaître (1994) that gave subarcsecond images in small patches within a 20 degree field.

The optical design of Wang, Moretto, Borra and Lemaître (1994) finds that $Coma5$, $Astm5$ are not negligible but we did not attempt to generate them at this time. These residual higher-order aberrations have lower amplitudes but the forces required for their correction are greater than those needed for the corrections that we have generated in this work. Furthermore, warping a mirror with those complex shapes will require more positions. The forces required for $Coma5$, $Astm5$ are not yet available but this will be the subject of a future investigation at which time we will measure the deformation limit of this active mirror.

In addition to obvious applications to the optical design of space instruments, groundbased telescopes could benefit from such coma-and astigmatism-corrected mirrors and gratings. Possibilities also exist in the design of focal instruments by off-axis optics.

**Table 3.** The aberration coefficients in wavelenght units for the following wavefronts measured from the active mirror:

(**A**) $Z = z_{33}$,
(**B**) $Z = z_{11} + z_{22} + z_{31}$,
(**C**) $Z = z_{11} + z_{22} + z_{31} + z_{33}$,
(**D**) $Z = z_{11} + z_{20} + z_{22} + z_{31} + z_{33} + z_{40}$.

| Aberration | Mode A | Mode B | Mode C | Mode D |
|---|---|---|---|---|
| $Tilt1$ | 0.227 | 0.586 | 0.684 | 0.875 |
| $Cv$ (Defocus) | 0.142 | 1.532 | 0.493 | 3.549 |
| $Astm3$ | 0.592 | 6.459 | 5.670 | 4.933 |
| $Coma3$ | 0.431 | 5.574 | 4.628 | 5.161 |
| $Sph3$ | 0.219 | 0.347 | -0.708 | -5.387 |
| $Comatri$ | 4.344 | 0.237 | 2.052 | 3.362 |

*Acknowledgements.* This research has been supported by the Natural Sciences and Engineering Research Council of Canada and partly supported by Université de Provence and by a grant from Ministère des Affaires Etrangères No 545-STE-01-AM (Collaboration Franco-Québecoise). G. Moretto was also supported by CAPES - Brazil and Collaboration Franco-Québecoise via CIES-France.

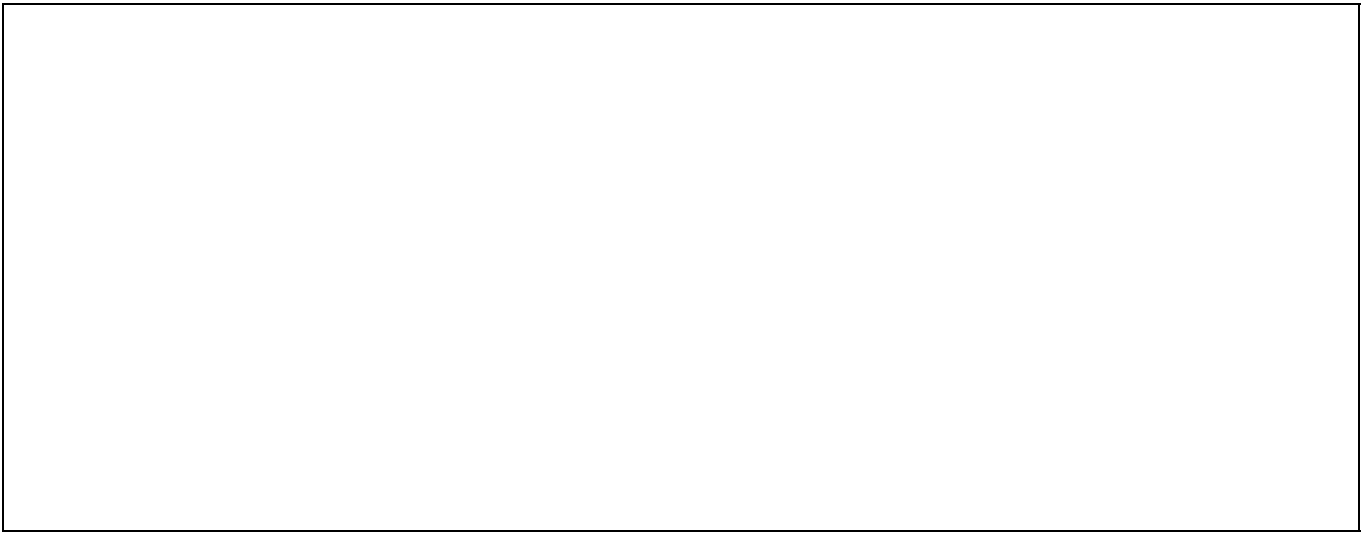

**Fig. 8.** Interferometric pattern for the superposed modes:

**(A)** $Z = z_{11} + z_{22} + z_{31}$,
**(B)** $Z = z_{11} + z_{22} + z_{31} + z_{33}$,
**(C)** $Z = z_{11} + z_{20} + z_{22} + z_{31} + z_{33} + z_{40}$.

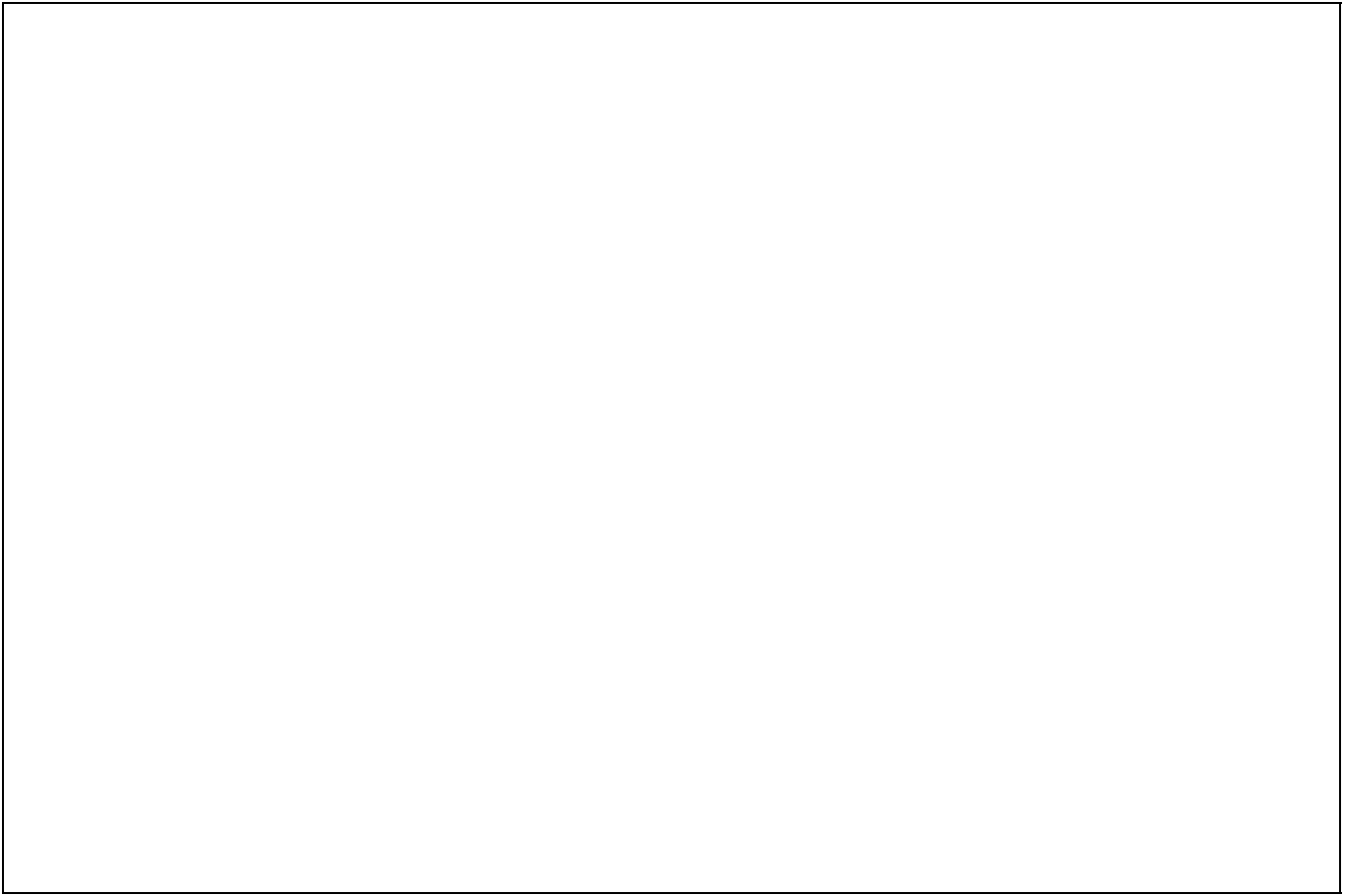

**Fig. 9.** The contour and isometric maps of the wavefronts:

**(A)** $Z = z_{33}$,
**(B)** $Z = z_{11} + z_{22} + z_{31}$,
**(C)** $Z = z_{11} + z_{22} + z_{31} + z_{33}$,
**(D)** $Z = z_{11} + z_{20} + z_{22} + z_{31} + z_{33} + z_{40}$.